%Paper: hep-th/9206071
%From: hamada@tkyvax.phys.s.u-tokyo.ac.jp (kenji hamada)
%Date: Thu, 18 Jun 92 18:44:58 +0900

\input phyzzx

\def\df{\varphi}
\def\gr{g_{\rho}}
\def\gs{g_{\sigma}}
\def\bg{{\hat g}}
\def\bgr{{\rm e}^{2\rho} {\hat g}}
\def\bgs{{\rm e}^{2\sigma} {\hat g}}
\def\blap{{\hat \Delta}}
\def\bcv{{\hat R}}
\def\vr{\delta \rho}
\def\vs{\delta \sigma}
\def\vg{\delta g}
\def\vp{\delta \df}
\def\vf{\delta f}
\def\dg{\hbox{$\sqrt{-g}$}}
\def\dbg{\hbox{$\sqrt{-{\hat g}}$}}
\def\e{{\rm e}}
\def\det{{\rm det}}
\def\vepsi{\varepsilon}

\REF\bkkm{D. Boulatov, V. Kazakov, I. Kostov and A. Migdal, Nucl. Phys.
         {\bf B275} (1986) 641;
         F. David, Nucl. Phys. {\bf B257[FS14]} (1985) 543;
         J. Ambj{\o}rn, B. Durhuus and J. Fr\"ohlich, Nucl. Phys. {\bf B257}
         (1985) 433.}
\REF\gm{D. Gross and A. Migdal, Phys. Rev. Lett. {\bf 64} (1990) 127;
       Nucl. Phys. {\bf B340} (1990) 333;
       M. Douglas and S. Shenker, Nucl. Phys. {\bf B335} (1990) 635;
       E. Brezin and V. Kazakov, Phys. Lett. {\bf 236} (1990) 144.}
\REF\kpz{V. Knizhnik, A. Polyakov and A. Zamolodchikov, Mod. Phys. Lett.
         {\bf A3} (1988) 819.}
\REF\dk{J. Distler and H. Kawai, Nucl. Phys. {\bf B321} (1989) 509;
        F. David, Mod. Phys. Lett. {\bf A3} (1988) 1651.}
\REF\s{N. Seiberg, Proc. of the 1990 Yukawa Int. Seminor, to appear in
        Prog. Theor. Phys. Supp. {\bf 102} (1990) 319.;
       J. Polchinski, Proc. of the String 1990, Texas A\&M, March 1990.}
\REF\p{J. Polchinski, Nucl. Phys. {\bf B357} (1991) 241;
        K. Hamada, Nucl. Phys. {\bf B365} (1991) 354; {\it Ward Identities
        of Liouville Gravity coupled to Minimal Conformal Matter}, preprint
        UT-Komaba 91-25, to appear in Prog. Theor. Phys. Supp.}
\REF\cghs{C. Callan, S. Giddings, J. Harvey and A. Strominger, Phys. Rev.
          {\bf D45} (1992) R1005.}
\REF\rst{J. Russo, L. Susskind and L. Thorlacius, {\it Black Hole Evaporation
         in 1+1 Dimensions}, Stanford University preprint, SU-ITP-92-4;
         T. Banks, A. Dabholkar, M. Douglas and M. O'Loughlin, {\it Are Horned
         Particles the Climax of Hawking Evaporation?} Rutgers University
         preprint, RU-91-54.}
\REF\w{S. Hawking, {\it Evaporation of Two Dimensional Black Hole}, Caltech
       preprint, CALT-68-\#1;
       L. Susskind and L. Thorlacius, {\it Hawking Radiation and Backreaction},
       Stanford preprint, SU-ITP-92-12;
       B. Birnir, S. Giddings, J. Harvey and A. Strominger, {\it Quantum
       Black Holes}, preprint UCSB-TH-92-08, EFI-92-16.}
\REF\t{A. Tomimatsu, {\it Evaporating Black Boles in Quantum Gravity},
       Nagoya University preprint.}
\REF\bd{N. Birrell and P. Davies, {\it Quantum Fields in Curved Space},
       Cambridge University Press.}
\REF\tih{P. Thomi, B. Isaak and P. Hajicek, Phys. Rev. {\bf D30} (1984) 1168;
         P. Hajicek, Phys. Rev. {\bf D30} (1984) 1178.}
\REF\str{A. Strominger, {\it Fadeev-Popov Ghosts and 1+1 Dimensional Black
        Hole Evaporation}, preprint UCSB-TH-92-18;
        S. deAlwis, {\it Quantization of a Theory of 2d Dilaton Gravity},
        preprint COLO-HEP-280;
        A. Bilal and C. Callan, {\it Liouville Models of Black Hole
        Evaporation}, preprint PUPT-1320.}

\pubnum{UT-Komaba 92-7}

\titlepage

\title{{\bf Quantum Theory of Dilaton Gravity in 1+1 Dimensions}}

\author{Ken-ji Hamada}

\address{Institute of Physics, University of Tokyo  \break
Komaba, Meguro-ku, Tokyo 153, Japan}

\abstract{We discuss the quantum theory of 1+1 dimensional dilaton gravity,
which is an interesting model with analogous
features to the spherically symmetric gravitational systems in 3+1 dimensions.
The functional measures over the metrics and the dilaton field are explicitly
evaluated and the diffeomorphism invariance is completely fixed in conformal
gauge by using the technique developed in the two dimensional quantum gravity.
We argue the relations to the ADM formalism. The physical state conditions
reduce to the usual Wheeler-DeWitt equations when the dilaton
$\df^2 ~ (=\e^{-2\phi}) $ is large enough compared with $\kappa =(N-51/2)/12$,
where $N $ is the number of matter fields. This corresponds to the large mass
limit in the black hole geometry. A singularity appears at
$\df^2 =\kappa (>0) $.
The final stage of the black hole evaporation corresponds to the region
$\df^2 \sim \kappa $, where the Liouville
term becomes important, which just comes from the measure of the metrics.
If $\kappa < 0 $, the singularity disappears.}

\endpage

\chapter{{\bf Introduction}}

    The quantization of gravity is a long standing issue in  theoretical
physics and a sufficient solution has not been avaiable for a long time. In
recent years, however, we got the complete solutions in the two dimensional
quantum gravity. The model was solved in the lattice gravity (or the matrix
model) method\refmark{\bkkm,\gm}
and the field theoretical (or Liouville gravity)
method\refmark{\kpz,\dk,\s,\p}
in which the functional integration over metrics is carried out
explicitly. Unfortunately the model is now solved only in the case that the
dynamical degrees of freedom are very few.

    A dynamical model of the gravitation in 1+1 dimensions has been proposed
by Callan et al.\refmark{\cghs}, which is called the dilaton gravity.
The features of the model are very
similar to the spherically symmetric gravitational system in 3+1 dimensions.
An advantage of this model is very simple to manage.
The several authors\refmark{\rst,\w} solved
the model and estimated the back-reaction of the Hawking radiation in the
semi-classical approximation.

     The quantum gravity becomes very important at the final stage of the
black hole evaporation.  As a quantization method of the gravitation,
there is the Arnowitt-Deser-Misner (ADM) formalism or
Wheeler-DeWitt approach. Recently Tomimatsu\refmark{\t}
derived the interesting
results for the spherically symmetric black hole. He showed that by solving the
Wheeler-Dewitt equations in the local mini-superspace  approximation near the
apparent horizon, the mass of the black hole $M $ decreases with the law
$<{\dot M}> \propto -M^{-2} $ inferred by Hawking. The simillar arguments can
be done in the
parallel way in the case of the 1+1 dimensional dilaton gravity.

    However, there are some problems in the ADM formalism, the issues of
measures and orderings. In fact, if we apply the ADM formalism to the
two dimensinal gravity without dilaton, the Hamiltonian and the momentum
constraints become trivial. The non-trivial contributions exactly come from
the functional measure over metrics.

    In this paper we consider the quantization of the 1+1 dimensional dilaton
gravity. Then we explicitly evaluate the contributions of measures.
We first argue the measure of the dilaton field in Sect.2. The dilaton field
is coupled to the curvature, so we have to estimate it carefully.
Following the procedure of David-Distler-Kawai (DDK)\refmark{\dk},
we determine the
measure of  metrics in confomal gauge in Sect.3.
{}From the gauge fixed action  the physical state conditions are derived in
Sect.4.  Then we compare with the
ADM formalism. Our physical state conditions reduce to
the usual Wheeler-DeWitt equations in the large $\varphi $ $(= {\rm
e}^{-\phi})$ limit, which corresponds to the large mass limit of the black hole
geometry.

\chapter{{\bf Evaluation of the measure of dilaton field}}

   The theory of 1+1 dimensional dilaton gravity is defined by the following
action
$$
\eqalign{
       &  I(g,\df,f)=I_D (g,\df)+I_M (g,f) ~,               \cr
       &  I_D (g,\df)= {2 \over \pi} \int d^2 x \dg
               (g^{\alpha \beta} \partial_{\alpha} \df \partial_{\beta} \df
               + \xi R_g \df^2 + \lambda^2 \df^2 ) ~,        \cr
       &  I_M (g,f) = -{1 \over 4\pi} \sum^N_{j=1} \int d^2 x \dg
           g^{\alpha \beta} \partial_{\alpha} f_j \partial_{\beta} f_j ~, \cr
        } \eqno\eq
$$
where $\df ={\rm e}^{-\phi} $ is the dilaton field and $f_j $'s are $N$ matter
fields. $\lambda $ is the cosmological constant and $\xi =1/4 $. $R_g $ is the
curvature of the metrics $g $. The classical equations of motion can be easily
solved and one obtains, for instance, the black hole geometry
$$
        \df^2 = {\rm e}^{-2\rho} = {M \over \lambda} -\lambda^2 x^+ x^- ~,
          f_j =0 ,
        \eqno\eq
$$
where $g_{\alpha \beta}={\rm e}^{2\rho} \eta_{\alpha \beta} $,
$\eta_{\alpha \beta} =(-1,1) $ and $x^{\pm} =x^0 \pm x^1 $. $M $ is the mass
of the black hole.

    The corresponding quantum theory is defined by
$$
    Z = \int {D_g(g) D_g(\df) D_g(f) \over Vol(Diff.) }
           {\rm e}^{iI(g,\df,f)} ~,
     \eqno\eq
$$
where $Vol(Diff.) $ is the gauge volume. The functional measures are defined
from the following norms
$$
\eqalign{
    &  \Vert \vg \Vert^2_g =
           \int d^2 x \dg g^{\alpha \gamma} g^{\beta \delta}
              \vg_{\alpha \beta} \vg_{\gamma \delta} ~,        \cr
    &  \Vert \vp \Vert^2_g =
           \int d^2 x \dg \vp \vp  ~,                           \cr
    &  \Vert \vf_j \Vert^2_g =
           \int d^2 x \dg \vf_j \vf_j  ~.                       \cr
          }   \eqno\eq
$$

    We first consider the functional integration over $\df $, which gives
the following determinant
$$
       \int D_g(\df) \e^{iI_D(g,\df)} = L [ \det_g D ]^{-1/2} ~,
          \eqno\eq
$$
where the operator $D $ is defined by
$$
       D= \Delta_g + \xi R_g + \lambda^2
           \eqno\eq
$$
and $L $ is a constant factor. The Laplacian is $\Delta = -\nabla^{\alpha}
\nabla_{\alpha} $. For a while we treat $\xi $ as a free parameter.
Now we decompose the metrics into the Weyl mode $\sigma $ and the background
metrics $\bg $ as $ g=\gs \equiv \bgs $. From now on we concentrate on the
$\sigma $-dependence of $ \det_{\gs} D $. Noting the relation
$ R_{\gs} = \e^{-2\sigma}(\bcv +2 \blap \sigma) $, one gets
$$
        D=\e^{-2\sigma} \blap
           + \xi \e^{-2\sigma} (\bcv +2\blap \sigma ) +\lambda^2 ~.
       \eqno\eq
$$
The variation of $D $ with respect to $\sigma $ is given by
$$
        \delta_{\sigma} D = -2\vs D +2\xi \e^{-2\sigma} \blap \vs
                           + 2\lambda^2 \vs ~.
          \eqno\eq
$$
Therefore the $\sigma $-dependent terms of $\det_{\gs} D $ are evaluated as
$$
\eqalign{
       \delta_{\sigma} {\rm log} \det_{\gs} D & =
              \int^{\infty}_{\vepsi} i dt Tr_{\gs}
                \bigl( \delta_{\sigma} D \e^{-it D} \bigr)   \cr
       & = -2 Tr_{\gs} (\vs \e^{-i\vepsi D} )
             + 2\xi Tr_{\gs} (\Delta \vs D^{-1} )            \cr
       & \qquad \qquad + 2\lambda^2 Tr_{\gs} (\vs D^{-1} ) ~,  \cr
         }  \eqno\eq
$$
where $\vepsi $ is a infinitesimal parameter to regularize divergences. And  we
also evaluate the integral by giving a small imaginary cosmological
constant as $\lambda^2 -i \vepsi^{\prime} $. The trace is
$$
      Tr_g (F) = \int d^2 x \dg <x \vert F \vert x>_g ~,
           \qquad <x \vert x^{\prime}>_g ={1 \over \dg }
                    \delta^2 (x-x^{\prime} ) ~.
      \eqno\eq
$$

     The first term of eq.(2.9) can be easily evaluated. The Kernel
$<x \vert \e^{-itD} \vert x>_g $ was already calculated
(see for instance ref.[11]).
The result
is
$$
\eqalign{
       <x \vert \e^{-itD} \vert x>_g
          &  = {i \over 4\pi} \e^{-i\lambda^2 t} {1 \over it}
               ( a_0 (x) +a_1 (x) it + a_2 (x) (it)^2 + \cdots ) ~,  \cr
          &                                                    \cr
         a_0 (x) & = 1 ~,                                       \cr
         a_1 (x) & = \biggl( {1 \over 6} -\xi \biggr) R_g  ~,   \cr
         a_2 (x) & = \biggl[ {1 \over 2} \biggl( {1 \over 6} -\xi \biggr)^2
                             + {1 \over 360} \biggr] R^2_g
                        - {1 \over 6} \biggl( {1 \over 5} -\xi \biggr)
                                \Delta_g R_g    ~,              \cr
                 & \cdots    ~.                                 \cr
         }    \eqno\eq
$$
{}From this we obtain
$$
\eqalign{
     -2 Tr_{\gs} ( \vs \e^{-i\vepsi D} )
           = & {1 \over 2\pi } \biggl( -{1 \over \vepsi} +i\lambda^2 \biggr)
                     \int d^2 x \hbox{$\sqrt{-\gs}$} \vs          \cr
           & \qquad -{i \over 2\pi } \biggl( {1 \over 6} -\xi \biggr)
                 \int d^2 x \hbox{$\sqrt{-\gs}$} R_{\gs} \vs +O(\vepsi)  \cr
         }   \eqno\eq
$$
The second and the third terms of (2.9) are not so easy, which are evaluated
only in the curvature expansion (2.11).

     However, we want to know only the Weyl mode dependence of the measure.
So we also  evaluate the following determinant
$$
      \int D_{\bg} (\df) \e^{iI_D (\gs, \df)}
           = L [ \det_{\bg} \e^{2\sigma} D ]^{-1/2}
           \equiv L [ \det_{\bg} {\hat D} ]^{-1/2}  ~,
      \eqno\eq
$$
and compare with the result (2.9),
where $D_{\bg} (\df) $ is defined by the norm (2.4) with $g = \bg $.
The variation of the operator
$ {\hat D} =\blap +\xi (\bcv +2\blap \sigma ) +\lambda^2 \e^{2\sigma} $
with respect to $\sigma $ is given by
$$
      \delta_{\sigma} {\hat D}
             = 2\xi \blap \vs +2 \lambda^2 \e^{2\sigma} \vs ~.
        \eqno\eq
$$
Thus the $\sigma $-dependent terms of $\det_{\bg} {\hat D} $ are
$$
\eqalign{
       \delta_{\sigma} {\rm log} \det_{\bg} {\hat D}
           & = 2\xi Tr_{\bg} ( \blap \vs {\hat D}^{-1} )
                + 2\lambda^2 Tr_{\bg} ( \e^{2\sigma} \vs {\hat D}^{-1} )  \cr
           & = 2\xi Tr_{\gs} ( \Delta_{\gs} \vs  D^{-1} )
                + 2\lambda^2 Tr_{\gs} ( \vs D^{-1} )  ~,         \cr
         }   \eqno\eq
$$
where we use the relation for the propagators
$$
      <x \vert {\hat D}^{-1} \vert x^{\prime} >_{\bg}
            = <x \vert D^{-1} \vert x^{\prime} >_{\gs}  ~.
         \eqno\eq
$$
The difference of the expressions (2.9) and (2.15) is what we want.
Combining eqs.(2.9), (2.15) and (2.12), we get
$$
\eqalign{
          &  \delta_{\sigma}  {\rm log} \det_{\gs} D
              -  \delta_{\sigma} {\rm log} \det_{\bg} {\hat D}      \cr
          & = -2 Tr_{\gs}(\vs \e^{-i\vepsi D} )                      \cr
          & = \delta_{\sigma} \biggl[ {1 \over 4\pi}
                  \biggl( -{1 \over \vepsi} +i\lambda^2 \biggr)
                   \int d^2 x \dbg \e^{2\sigma}
                  - {i \over 12\pi } (1-6\xi)
                   \int d^2 x \dbg (\sigma \blap \sigma +\bcv \sigma)
                    \biggr]  ~.                                    \cr
        }   \eqno\eq
$$
This means that the following relation is realized,
$$
\eqalign{
       &  \int D_{\bgs} (\df) \e^{iI_D (\bgs, \df)}                  \cr
       & \quad = \exp \biggl[ {1 \over 8\pi}
                   \biggl( {1 \over \vepsi} -i \lambda^2 \biggr)
                   \int d^2 x \dbg \e^{2\sigma}
                   +i{c_{\df} \over 12\pi } S_L (\sigma, \bg) \biggr]
                 \int D_{\bg} (\df) \e^{iI_D (\bgs, \df)}  ~,       \cr
        }    \eqno\eq
$$
where $S_L $ is the Liouville action
$$
        S_L (\sigma, \bg) = {1 \over 2}
                  \int d^2 x \dbg ( \bg^{\alpha \beta}
                   \partial_{\alpha} \sigma \partial_{\beta} \sigma
                   + \bcv \sigma )
          \eqno\eq
$$
and $ c_{\df} $ is defined by
$$
        c_{\df} = 1-6\xi ~ .
           \eqno\eq
$$
At $ \sigma =0 $, the both sides of the expression (2.18) should be equal.
So the divergent term of the form $\Lambda \int d^2 x \dbg \e^{2\sigma} $
should be renormalized properly.

\chapter{\bf{Conformal gauge fixing}}

     The functional measure over metrics $g_{\alpha \beta} $ is defined by
the norm (2.4). We decompose integrations over metrics $g_{\alpha \beta} $
into integrations over vector fields $v_{\alpha} $ which generate
infinitesimal diffeomorphisms and an integration over the Weyl mode $\rho $,
$g=\e^{2\rho} \bg $. The integrations over $v_{\alpha} $  cancel
 out the gauge volume. The Jacobian can be represented by a functional
integral over ghosts $b $, $c $. Thus the partition function (2.3) becomes
$$
     Z = \int D_{\gr}(\rho) D_{\gr}(\df) D_{\gr}(f) D_{\gr}(b) D_{\gr}(c)
           \e^{iI_D (\gr,\df)+iI_M (\gr,f)+iI_{gh}(\gr,b,c) }    ~,
      \eqno\eq
$$
where $\gr \equiv \e^{\rho} \bg $ and $I_{gh} $ is the well-known ghost action.
% which is given by
%$$
%     I_{gh} (g,b,c) = -{1 \over 2\pi}
%        \int d^2 x \dg (b_{++} \nabla^+ c^+ + b_{--} \nabla^- c^- )  ~.
%     \eqno\eq
%$$
The measure $D_{\gr} (\rho) $ is defined from the norm (2.4) by
$$
      \Vert \vr \Vert^2_{\gr}
          = \int d^2 x \hbox{$\sqrt{-\gr}$} (\vr)^2
          = \int d^2 x \dbg \e^{2\rho} (\vr)^2  ~.
      \eqno\eq
$$
This measure is very inconvenient because the norm is not invariant under the
shift $\rho \rightarrow \rho +h $. According to DDK\refmark{\dk},
we rewrite the measure
$D_{\gr} (\rho) $ into the covenient one $D_{\bg} (\rho) $ which is defined by
the norm
$$
      \Vert \vr \Vert^2_{\bg} = \int d^2 x \dbg (\vr)^2   ~.
        \eqno\eq
$$
It is invariant under the shift of $\rho$. To rewrite the measure we need the
Jacobian, which is assumed from the experience of the two dimensional quantum
gravity  as
$$
       D_{\gr} (\rho) = D_{\bg} (\rho)
            \exp \biggl[ i {A \over 12\pi} S_L (\rho, \bg) \biggr] ~.
       \eqno\eq
$$
The action $S_L (\rho, \bg) $ is defined by eq.(2.19) with $\sigma =\rho $.
The parameter $A$ is determined by the consistency.

    The transformation property of the measure $D_{\gr} (\df) $ under Weyl
rescalings is calculated in the
previous section. For the measures of the matters and the ghost fields
the transformation property is well-known  and given by
$$
     D_{\gr}(f) D_{\gr}(b) D_{\gr}(c)
          = \exp \biggl[ i{N-26 \over 12\pi} S_L (\rho, \bg) \biggr]
            D_{\bg}(f) D_{\bg}(b) D_{\bg}(c)  ~.
     \eqno\eq
$$
Here note that the actions of the matters and the ghost fields are invariant
under Weyl rescalings. Using these expressions  we can rewrite the
partition function as
$$
\eqalign{
       \int D_{\bg}(\rho) & D_{\bg}(\df) D_{\bg}(f) D_{\bg}(b) D_{\bg}(c)
          \exp \biggl[ i {B \over 12\pi} S_L (\rho, \bg)             \cr
          &  + iI_D (\e^{2\rho} \bg,\df) +iI_M (\bg,f)
               +iI_{gh}(\bg,b,c) \biggr] ~,                          \cr
        }   \eqno\eq
$$
where
$$
         B=A+c_{\df} +N-26  ~.
             \eqno\eq
$$
Note that, when we rewrite the partition function (2.3) into this form, the
divergent terms appear, which always have the form proportional to
$ \int d^2 x \dg $. To renormalize the divergences we intoroduce a bare
term $\mu_0  \int d^2 x \dg $ and cancel out these by adjusting the bare
constant $\mu_0 $. The renormalized constant is set to zero. In the following
we do not consider this type of divergence.

   To determine the parameter $A $ we use the fact that the original theory
depends only on the metrics $\gr = \bgr $. This means that the theory should
be invariant under the simultaneous shift
$$
        \rho \rightarrow \rho - \sigma ~, \qquad \bg \rightarrow \bgs ~.
          \eqno\eq
$$
Applying the shift to the partition function, we obtain
$$
\eqalign{
     \int D_{\bgs}(\rho) & D_{\bgs}(\df) D_{\bgs}(f) D_{\bgs}(b) D_{\bgs}(c)
         \exp \biggl[ i {B \over 12\pi} S_L (\rho-\sigma, \bgs)           \cr
          &  + iI_D (\e^{2\rho} \bg,\df) +iI_M (\bg,f)
               +iI_{gh}(\bg,b,c) \biggr] ~,                          \cr
        }   \eqno\eq
$$
where we use the fact that the measure $D_{\bgs}(\rho) $ is invariant under
the shift of $\rho $.  This expression should return to the original
form (3.6).

    Let us argue changing the measures on the metrics $\gs \equiv \bgs $ into
the measures on the background metrics $\bg $. It is easily done for
the measures of the matters and the ghost fields  by using the relation (3.5)
(by replacing $\rho $ with $\sigma$ ). The Weyl transformation property of
the measure of $\df $
is given, from the previous calculation, by
$$
        \int D_{\bgs} (\df) \e^{iI_D (\bgr,\df)}
           = \exp \biggl[ i{c_{\df} \over 12\pi} S_L (\sigma, \bg) \biggr]
               \int D_{\bg}(\df) \e^{iI_D (\bgr, \df)}  ~.
        \eqno\eq
$$
In Sect.2 we discussed the case of $\rho =\sigma$. So we must show that the
expression (3.10) holds good in the case of $\rho \not= \sigma $.
It is realized by
seeing the $\rho $-dependence of the both sides of eq.(3.10). The left hand
side gives the determinant $L [\det_{\gs} {\tilde D} ]^{-1/2} $, where
${\tilde D} = \e^{-2\sigma}\e^{2\rho}D_{\gr} $ and
$ D_{\gr} = \Delta_{\gr} +\xi R_{\gr} +\lambda^2 $. While the functional
integration of the right hand side gives $ L[\det_{\bg} {\hat D}_{\rho}
]^{-1/2} $
where ${\hat D}_{\rho} = \e^{2\rho} D_{\gr} =\blap +\xi (\bcv +2\blap \rho)
+\lambda^2 \e^{2\rho} $. Then we obtain the following relation with respect
to the $\rho $-dependence,
$$
\eqalign{
       \delta_{\rho} \log \det_{\gs} {\tilde D}
          & = \int d^2 x \hbox{$\sqrt{-\gs}$}
               (2\xi \e^{-2\sigma} \blap \vr
                   +2\lambda^2 \e^{2(\rho-\sigma)} \vr)
                <x \vert {\tilde D}^{-1} \vert x>_{\gs}       \cr
          & = \int d^2 x \dbg
               (2\xi \blap \vr +2\lambda^2 \e^{2\rho} \vr)
                <x \vert {\hat D}^{-1}_{\rho} \vert x>_{\bg}         \cr
          & = \delta_{\rho} \log \det_{\bg} {\hat D}_{\rho} ~.       \cr
         }   \eqno\eq
$$
This means that the difference of the functional integrations in eq.(3.10)
depends only on $\sigma$, which is determined from the previous result
at $\rho =\sigma$.

    By using the expressions (3.5) and (3.10) and also the relation for the
Liouville action
$$
       S_L (\rho-\sigma, \bgs) = S_L (\rho,\bg)-S_L (\sigma,\bg) ~,
        \eqno\eq
$$
the partition function (3.9) reduces to the following form
$$
\eqalign{
     & \exp \biggl[ -i {A \over 12\pi}S_L (\sigma,\bg) \biggr]
      \int D_{\bgs}(\rho) D_{\bg}(\df) D_{\bg}(f) D_{\bg}(b) D_{\bg}(c)  \cr
     & \quad \times \exp \biggl[ i{B \over 12\pi} S_L (\rho,\bg)
        +iI_D (\bgr,\df) +iI_M (\bg,f) +iI_{gh}(\bg,b,c) \biggr] ~.      \cr
        }    \eqno\eq
$$

      Finally consider changing the measure $D_{\bgs}(\rho) $ into
$D_{\bg} (\rho)$. As the measure is invariant under the shift of $\rho $,
we can replace $\rho $ with $\rho^{\prime} = \rho +h $, where
$ h= (1/2)\blap^{-1} \big[ \bcv +(96\xi /B)\blap \df^2 \big]$. Then the
$\rho^{\prime} $ integration becomes that of the single free
boson\footnote\dagger{Here we do not care the cosmological constant term. From
the experience of the 2 dimensional quantum gravity, it probably does not
contribute to the value of the central charge.}.
%%%%
Since the shift $h $ is independent of $\rho $ and  $\sigma $, the Weyl
dependence of the $\rho $-measure  corresponds to the case of central charge
$ 1 $. Therefore, if we set
$$
          A=1   ~,
               \eqno\eq
$$
the Liouville action $S_L (\sigma,\bg) $ completely cancels out and the
partition function reduces to the original form.

    Setting $\xi =1/4 $, we finally get the conformal gauge fixed action
of the 1+1 dimensional dilaton gravity
$$
\eqalign{
   {\hat I} = & {1 \over 2\pi} \int d^2 x \dbg \Bigl[
    4{\hat g}^{\alpha \beta} \partial_{\alpha} \df \partial_{\beta} \df
    +4{\hat g}^{\alpha \beta} \df \partial_{\alpha} \df \partial_{\beta} \rho
       + \bcv \df^2 +4 \lambda^2 \df^2 \e^{2\rho}          \cr
      & +{N-51/2 \over 12} (
         {\hat g}^{\alpha \beta} \partial_{\alpha} \rho \partial_{\beta} \rho
         + \bcv \rho )
       -{1 \over 2} \sum^N_{j=1}
         {\hat g}^{\alpha \beta} \partial_{\alpha} f_j \partial_{\beta} f_j
       \Bigr] +I_{gh}(\bg,b,c) ~.                            \cr
        }  \eqno\eq
$$

     We showed that the theory is invariant under the simultaneous
shift (3.8).  Since the measure of $\rho $ defined by (3.3) is invariant
under local shifts of $\rho $, the theory is invariant under conformal
changes of the background metrics $\bg $.

\chapter{{\bf Physical state conditions}}

     Now we carry out the canonical quantization of the gauge-fixed 1+1
dimensional dilaton gravity. Since the functional measures are defined
on the background metrics $\bg_{\alpha \beta} $,  we can set the canonical
commutation relations in
usual way. Here we choose the flat metric $\eta_{\alpha \beta} $ as the
background metric $\bg_{\alpha \beta}$. Then the canonically conjugate
momentums for $\df $, $\rho $ and $ f_j $ are given by
$$
\eqalign{
     \Pi_{\df} &= -{4 \over \pi} {\dot \df}
                       -{2 \over \pi} \df {\dot \rho}  ~,      \cr
     \Pi_{\rho}&= -{N-51/2 \over 12\pi} {\dot \rho}
                       -{2 \over \pi} \df {\dot \df} ~,        \cr
     \Pi_{f_j} &= {1 \over 2\pi}  {\dot f}_j  ~,               \cr
        } \eqno\eq
$$
where the dot and the prime stand for the derivative with respect to the
time and space coordinate respectively.

    The physical state condition is defined from the independence of how to
choose the background metrics $\bg_{\alpha \beta} $.
So we set\footnote\dagger{In the case of the two dimensional gravity without
dilaton, this corresponds to the heighest state conditions of the Virasoro
algebra: $L_n \vert phys> ={\bar L}_n \vert phys> =0 $ for $n \geq 0 $, which
means that for arbitrary non-singular function $\epsilon $ and
$\epsilon^{\prime} $, $\int \epsilon {\hat T}_{00} \vert phys> =
\int \epsilon^{\prime} {\hat T}_{01} \vert phys> = 0 $.}
%%%%%
$$
       \langle {\delta {\hat I} \over \delta \bg^{\alpha \beta}} \rangle =0
           \eqno\eq
$$
or
$$
       {\hat T}_{00} \Psi ={\hat T}_{01} \Psi =0 ~,
            \eqno\eq
$$
where the energy-momentum tensor ${\hat T}_{\alpha \beta} $ is defined by
${\hat T}_{\alpha \beta}=-{2 \over \dbg}{\delta {\hat I} \over \delta
\bg^{\alpha\beta}}$.
$\Psi $ is a physical state. The condition for ${\hat T}_{11} $ reduces to the
one for $ {\hat T}_{00} $ by using the $\rho $-equation of motion.
In the flat background metric, the physical state conditions (4.3)
become\footnote\ddagger{For the zero-mode part, we maybe need the further
arguments.}
%%%%%%%%%
$$
\eqalign{
       \biggl[ {\pi/2 \over \df^2 -\kappa}
           & \Bigl( \Pi^2_{\rho} -\df \Pi_{\df} \Pi_{\rho}
                  + {\kappa \over 4} \Pi^2_{\df} \Bigr)
           + {2 \over \pi}  \bigl(  \df \df^{\prime\prime}
                 -\df \df^{\prime} \rho^{\prime}
                 -\lambda^2 \df^2 \e^{2\rho} \bigr)                  \cr
           & -{\kappa \over 2\pi}
               \bigl( \rho^{\prime 2} -2\rho^{\prime\prime} \bigr)
           +\sum^N_{j=1}  \Bigl( \pi \Pi^2_{f_j}
                   + {1 \over 4\pi} f^{\prime 2}_j \Bigr)
                     \biggr]  \Psi =0                                 \cr
         }    \eqno\eq
$$
and
$$
      \bigl( \df^{\prime} \Pi_{\df} + \rho^{\prime} \Pi_{\rho}
              -\Pi^{\prime}_{\rho} +\sum^N_{j=1} \Pi_{f_j} f^{\prime}_j
                   \bigr) \Psi =0    ~,
       \eqno\eq
$$
where $\kappa $ is defined by
$$
        \kappa = {N-51/2 \over 12} ~.
              \eqno\eq
$$

    If we rewrite the canonical momentums as the differential operators
$$
     \Pi_{\rho} = {\delta \over i\delta \rho} ~, \quad
     \Pi_{\df} = {\delta \over i\delta \df}   ~, \quad
     \Pi_{f_j} = {\delta \over i\delta f_j}   ~,
        \eqno\eq
$$
the eqs.(4.4) and (4.5) give the differential equations similar to the
Wheeler-DeWitt  equations\footnote\star{See for example ref.[12],  in which the
spherically
symmetric gravitational system of 3+1 dimensions is
discussed. Application to the 1+1 dimensional dilaton gravity is
straightforward.}.
%%%%%%%
The difference between  the usual Wheeler-DeWitt equations and ours is
just the Liouville term.
If $ \kappa > 0 $, there is a singularity at finite $\df^2 =\kappa $. Our
physical state conditions
reduce to the
usual form of the Wheeler-DeWitt equations in the limit $\df^2 \gg \kappa $. In
the black hole geometry this means that the mass of the black hole $M $ is
large enough  compared to $\lambda \kappa $ or the thinking region is far from
the singularity. So the usual Wheeler-DeWitt equations seem to be
correct in the semi-classical region. The final stage of the black hole
evaporation corresponds to the region $\df^2 \sim \kappa $,
where the Liouville term becomes important.

    The region $\kappa > \df^2 >0 $ is called the Liouville region, in which
the sign of the kinetic term of eq.(4.4) changes.  The existence of the
Liouville region is mysterious.

     $\kappa =0 $ is special. In this case the Liouville action disappears
and the physical state conditions reduce to the usual form of Wheeler-DeWitt
equations.  If $\kappa <0 $, the situation drastically changes. In this
region the singularity  disappears.

{\bf Note added}: After completing the calculation of the gauge fixing, we
received the preprints [13], in which the value of $c_{\df} $ is different of
ours. It seems that their calculation is esentially in the case of $\xi =0 $.

\ack{The author would like to thank T. Yoneya for valuable discussions.
This work is supported in part by Soryuushi Shogakukai.}

\endpage

\refout

\bye